\documentclass[preprint,12pt]{elsarticle}

\usepackage{graphicx}
\usepackage{amsmath,amssymb,amsfonts}
\usepackage{color}
\usepackage{algorithmic}
\usepackage{graphicx}
\usepackage{textcomp}
\usepackage{comment,color}
\usepackage{enumitem}
\usepackage{epsfig}
\usepackage{caption}
\usepackage{verbatim,bm,multicol,blindtext,subfigure}
\usepackage{epstopdf,array,mathtools,url}

\usepackage{lineno}

\journal{Elsevier}

\newcolumntype{L}{>{\centering\arraybackslash}m{3cm}}
\DeclarePairedDelimiter\norm{\lVert}{\rVert}%
\makeatletter
\let\oldabs\abs
\def\abs{\@ifstar{\oldabs}{\oldabs*}}

\let\oldnorm\norm
\def\norm{\@ifstar{\oldnorm}{\oldnorm*}}
\makeatother
\begin{document}

\begin{frontmatter}


\title{Wavefield solutions from machine learned functions}



\author[1]{Tariq Alkhalifah}
\ead{tariq.alkhalifah@kaust.edu.sa}
\address[1]{Physical Sciences and Engineering Division, King
Abdullah University of Science and Technology, Thuwal 23955, Saudi Arabia.}

\author[1]{Chao Song}

\author[2]{Umair bin Waheed}
\address[2]{Department of Geosciences, King Fahd University of Petroleum and Minerals, Dhahran 31261,  Saudi Arabia.}

%\cormark[1]

\author[3]{Qi Hao}
\address[3]{Center for Integrated Petroleum Research, King Fahd University of Petroleum and Minerals, Dhahran 31261,  Saudi Arabia.}

\begin{abstract}
Solving the wave equation is one of the most (if not the most) fundamental problems we face as we try to illuminate the Earth using recorded seismic data.	
The Helmholtz equation provides wavefield solutions that are dimensionally reduced, per frequency, compared to the time domain,
which is useful for many applications, like full waveform inversion (FWI). However, our ability to attain such wavefield solutions depends often on the size of the model and the complexity of the wave equation.
Thus, we use here a recently introduced framework based on neural networks to predict functional solutions through setting the underlying physical equation as a loss function
to optimize the neural network parameters. 
For an input given by a location in the model space, the network learns to predict the wavefield value at that location, and its partial derivatives using a concept referred to as automatic differentiation, to fit, in our case, a form of the Helmholtz equation. We specifically seek the solution of the scattered wavefield considering a simple homogeneous background model that allows for analytical solutions of the background wavefield. 
Providing the neural network (NN) a reasonable number of random points from the model space will ultimately
train a fully connected deep NN to predict the scattered wavefield function. The size of the network depends mainly on the complexity of the desired wavefield, with such complexity increasing with increasing frequency and increasing model complexity. However, smaller networks can provide smoother wavefields that might be useful for inversion applications. Preliminary tests on a two-box-shaped scatterer model with a source in the middle, as well as, the Marmousi  model with a source on the surface demonstrate the potential of the NN for this application. Additional tests on a 3D model demonstrate the potential versatility of the approach. 
\end{abstract}


\end{frontmatter}

%\linenumbers

\section{Introduction}
\label{section1}

A fundamental part of using surface seismic recorded data to illuminate the Earth is solving the wave equation \cite{claerbout1985imaging}. Solving the wave equation numerically constitutes 
the majority of the computational cost and complexity in
applications like seismic modeling, imaging, and waveform inversion. Time-domain solutions of the wave equation dominate seismic applications as they are often efficient and comply with our natural understanding of wave evolution~\cite{alterman1968propagation,richards1980quantitative}. However, frequency-domain solutions, providing a reduction in dimensionality, recently gained
additional attention with the rise of waveform inversion \cite{pratt1999seismic,sirgue2004efficient}. Such solutions are obtained by inverting the stiffness matrix of the Helmholtz wave equation. However, the cost and complexity of such a matrix inversion are intolerable as the model size increases, like for high frequencies or 3D applications \cite{clement1990conjugate}, or the wave equation is complex, like those in anisotropic media \cite{wu2018efficient}. This led \cite{sirgue20083d}  to suggest using time-domain modelling to obtain wavefields in the frequency domain for waveform inversion applications. However, such solutions are vulnerable to dispersion and stability errors \cite{wu2018highly}.

In recent years, researchers in our field have utilized machine learning algorithms to predict everything from fault locations to horizons to salt boundaries to facies classification, as well as, velocity models \cite{roth1994neural,wrona2018seismic,araya2019deep,holm2020linear}. Whether supervised or semi or unsupervised training,
neural networks have shown incredible flexibility in adapting to various geophysical tasks. Supervised learning was instrumental in predicting low frequencies to help full waveform inversion (FWI) converge to an accurate solution \cite{ovcharenko2019deep}. Deep learning was also utilized to develop a priori models from well information to be used in FWI \cite{mosser2018stochastic,zhang2019regularized}.
Even wave propagation and wave equation solutions were facilitated using deep neural networks \cite{sorteberg2018approximating,hughes2019wave}.

Within the framework of utilizing deep neural networks as universal function approximators \cite[]{liu1989limited} 
and under the banner of physics-informed neural networks (PINN), 
\cite{raissi2019physics} demonstrated the network's flexibility in learning how to
extract desired functional solutions to nonlinear partial differential equations, utilizing the concept of automatic differentiation \cite{baydin2018automatic}. PINN has found considerable traction in
solving partial differential equations (linear and nonlinear ones) ranging from cardiac activation mapping \cite{sahli2020physics} to steady-state Navier-Stokes
equation \cite{dwivedi2021distributed}.
Even with the framework of one dimensional wave propagation, PINN was utilized to establish flexible domain solutions of the wave equation \cite{kissas2020machine}.
In all of these applications, the predicted solutions were smooth, which is a requirement of NN as a universal function approximator \cite{pinkus1999approximation}. Wavefields are generally smooth, but they are often more complex in nature
than other physical phenomena. The complexity of the wavefield increases at the source, as it represents a singularity in the solution. Thus, to use ML to predict wavefield solutions will require larger neural networks, which will eventually require larger computational resources. It will also require, like most numerical methods, careful sampling of the source region.
As a result, \cite{song2021solving} suggested that we seek such NN functions for the scattered wavefield instead of the full wavefield. Specifically, to avoid the need for adaptive training points for the neural network (NN) to handle the expected source singularity bias, 
they solve for the scattered wavefield in the frequency domain, and thus, utilize the corresponding Lippmann–Schwinger equation as the loss function to train a deep fully connected neural network
with inputs given by (randomly chosen) points in space (within the domain of interest) and outputs given by the complex scattered wavefield at these points. In their implementation, they focus on the application of their method on anisotropic media. Our objective in this paper is to evaluate
what exactly an NN can predict of the wavefield solution, especially at a reasonable cost that can be utilized in practical applications.

Thus, here, we focus on the role of the model size, the solver, and frequency of the wavefield in predicting such scattered wavefield solutions. We will first
compare solutions for the Helmholtz equation \cite{mcfall2009artificial} to those obtained for the scattered version of the Helmholtz equation for the same network size and hyperparameters. We will compare solutions at two different frequencies to assess the ability of the NN model in handing higher frequencies. This is followed by investigating the role of the NN model size in smoothing the wavefield solutions
by evaluating the corresponding velocity for the predicted wavefields.
We test the performance of the NN on a two box-shaped scatterer model,
as well as, the Marmousi model, and in the process show the sensitivity of the approach to model size and frequency.  Further testing on 3D will hopefully demonstrate the role of the NN parameters optimizer.

\section{The Helmholtz equation}

The wave equation is often solved in the time domain, and such solutions are attained by extrapolating the wavefield in time simulating what happens in nature \cite{richards1980quantitative}.
Wavefields in the time domain, however, are large, as they are given by a four dimensional function in 3D media or a three dimensional function in 2D media for a given source.
In addition, the time axis, often, requires fine sampling to avoid aliasing, and an even finer sampling of time is required to avoid instability when solving the wave equation using finite-difference methods \cite{courant1928partial}.

As a result of its linear nature, the wave equation can be easily formulated in many domains, including the very useful frequency domain.
In this case, the resulting Helmholtz equation can be solved per frequency, with no requirements on frequency sampling, admitting a reduction in dimensionality of the wavefield solution.
The Helmholtz equation in an acoustic, isotropic, constant density medium, described by the velocity, $v$, is given by:
\begin{equation}
\left(\nabla^2+ k^2 \right) u (\mathbf{x})=f(\mathbf{x}), \,\,\, where  \,\,\,  k = \frac{\omega}{v}.
\label{eqn:eq1}
\end{equation}
In this case, 
the solution of such an equation is a complex wavefield, $u=\{u_r, u_i\}$, defined in the Euclidean space, 
with ${\bf x}= \{x, y, z\}$, and a function of the angular frequency, $\omega$. 
As a result, our time-domain solution is nothing but a superposition of frequency-domain solutions (inverse Fourier transform).

The point source nature of the source function, $f$, admits a singularity in the wavefield solution at the point source location. Such a singularity often causes
inaccuracies in numerical solutions of the Helmholtz equation near the source. As suggested by \cite{song2021solving}, such a limitation can be addressed by solving the Lippmann–Schwinger form of the wave equation \cite{lippmann1950variational}, instead. This equation is exact as we do not apply the Born
approximation.
Thus, to somewhat mitigate the source singularity, we solve for the scattered wavefield, $\delta u=u-u_0$, where $u_0$ is the background wavefield satisfying 
the same wave equation (equation~\ref{eqn:eq1}) for the background velocity $v_0$.
Defining the velocity model perturbation, $\delta m=\frac{1}{v^2}-\frac{1}{v_0^2}$, the scattered wavefield satisfies 
\begin{equation}
	\left(\nabla^{2} + \frac{\omega^2}{v^2} \right) \delta u = -\omega^{2 }\delta m \, u_{0}.
	\label{eqn:eq2}
\end{equation}
For the scattered wavefield, the source function is no longer confined in space, like the point source. It now depends on the perturbation model, which may extend the full space domain.
To allow for efficient evaluation of the background wavefield, we consider the background velocity, $v_0$, to be constant. For marine acquisition, we may choose this constant velocity to equal the water velocity to reduce the effect of the 
source singularity even further. The wavefield in an acoustic isotropic medium in 3D for a constant velocity and a point source located at ${\bf x_s}$, is given by:
\begin{eqnarray}
u_{0}({\bf x}) = \frac{e^{i \frac{\omega}{v_0}|{\bf x}-{\bf x_s}|}}{ 4 \pi |{\bf x}-{\bf x_s}|},
\label{eqn:eq7a}
\end{eqnarray}
where $i$ is the imaginary identity.
For 2D applications, the solution for acoustic isotropic media is given by
\begin{eqnarray}
u_{0}({\bf X}) = \frac{i}{4} H_0^{(2)}(\frac{\omega}{v_0}|{\bf x}-{\bf x_s}|),
\label{eqn:eq7b}
\end{eqnarray}
where $H_0^{(2)}$ is the zero-order Hankel function of the second kind, here ${\bf x}= \{x, z\}$~\cite{richards1980quantitative}.

Solving for the scattered wavefield will allow us later to utilize random samples of the space domain to train the neural network to provide the functional solution representing the scattered wavefield in the frequency domain.
The analytical solution for the background wavefield allows us to evaluate the wavefield instantly at any random point in the domain of interest. 
Next, we will see exactly how these formulations help the training of a functional neural network (NN).

\section{The neural network solution}

\begin{figure}[ht]
\begin{center}
\includegraphics[width=0.85\textwidth]{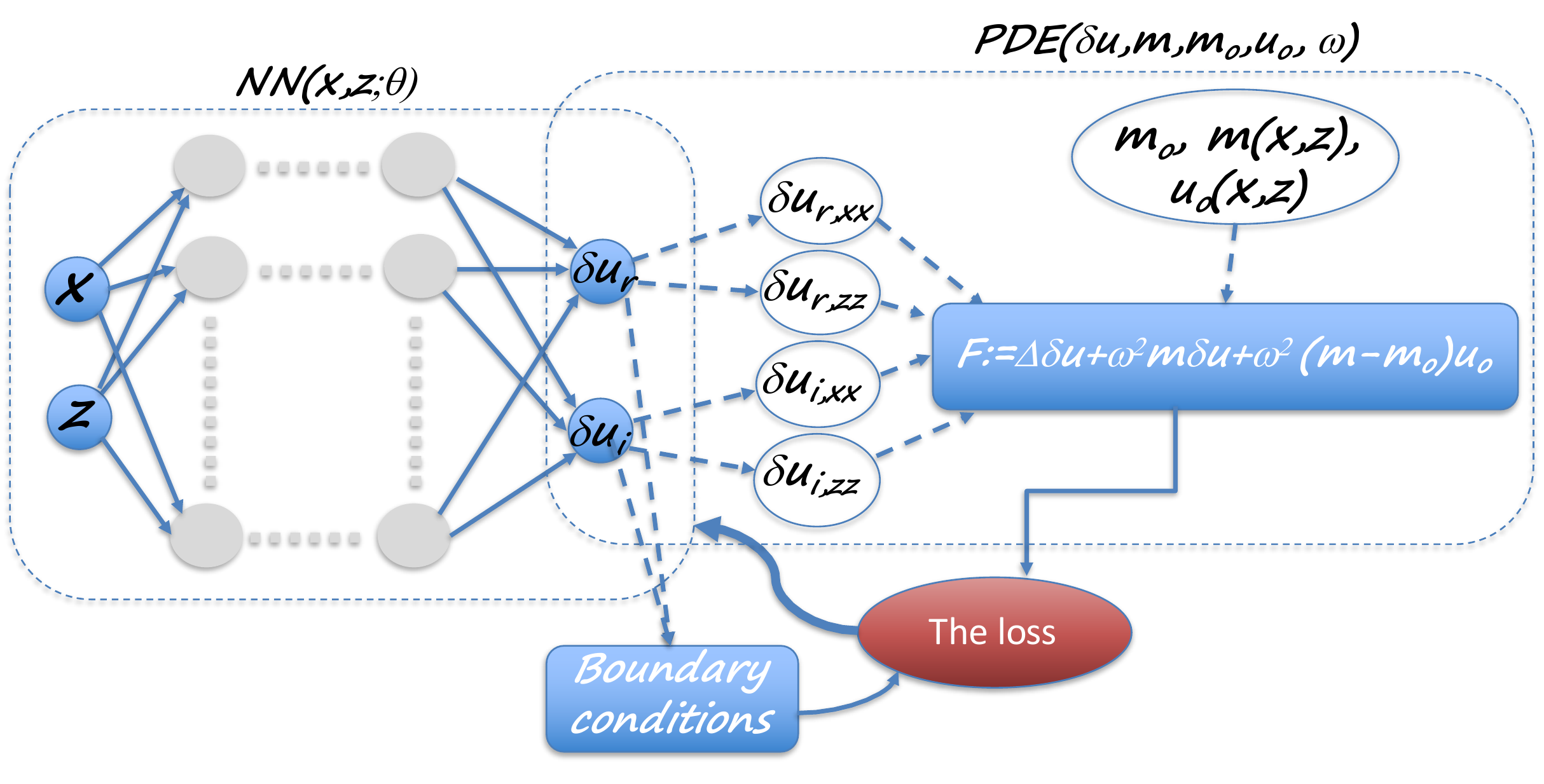}
\end{center}
\caption{
The neural network architecture (left side dashed box) with inputs ($x$, $z$) and outputs given by the real and imaginary parts of the (scattered) wavefield. 
The network is trained using a loss function given by the scattered wave equation (Right side dashed box), in which the Laplacian components ($\delta u_{r,xx}.\delta u_{r,zz},\delta u_{i,xx},\delta u_{i,zz}$) are
evaluated using automatic differentiation of the NN. The loss function can be supported by boundary conditions.
}%
\label{fig:diagramPINN}
\end{figure}

Based on the physics-informed neural network (PINN) framework introduced by \cite{raissi2019physics}, we utilize a neural network architecture using fully connected layers to approximate a function. This function is the scattered wavefield solution of equation~\ref{eqn:eq2}. 
\cite{hornik1989multilayer} have shown the ability of neural networks in approximating functions that are smooth, like what we would expect from solving the wave equation.
The input to the network, like a function, is a location in space, given in 2D by $x$ and $z$ coordinate values, and in 3D by $x$, $y$, and $z$  coordinate values. 
The output of the network consists of the real and imaginary values of the complex scattered wavefield at the input location. Figure~\ref{fig:diagramPINN} shows, in detail, the PINN concept for our application. We use the network to evaluate the wavefield and its second-order partial derivatives in $x$ and $z$, which is needed to evaluate the Laplacian operator and the loss function. Thus, to train the network, with equation~\ref{eqn:eq2}, we use the following loss function:
\begin{eqnarray}
f=\frac{1}{N}\sum_{j=1}^{N}  && \left | \omega ^{2}m^{(j)}\delta u_r^{(j)}+\nabla^{2}\delta u_r^{(j)}+\omega ^{2}\delta m^{(j)} u_{r0}^{(j)} \right |_{2}^{2} + \nonumber \\
										    &&  \left | \omega ^{2}m^{(j)}\delta u_i^{(j)}+\nabla^{2}\delta u_i^{(j)}+\omega ^{2}\delta m^{(j)} u_{i0}^{(j)} \right |_{2}^{2},
\label{eqn:eq3}
\end{eqnarray}
where $N$ is the number of training samples, and $j$ is the training sample index. The two terms in the loss function correspond to the losses for the real ($\delta u_r$) and imaginary ($\delta u_i$) parts of the scattered wavefield,
using the real ($u_{r0}$) and imaginary ($u_{i0}$) 
parts of the background wavefield. For the loss function, we chose the background model to be simple enough (homogeneous) so that the background wavefield can be evaluated analytically on the fly.
The details of the fully connected deep network will be shared in the examples. The activation function between layers, other than the last layer, in all the examples is an
inverse tangent. The last hidden layer connected to the output layer is linear. We chose to optimize the loss function using an Adam optimizer followed by limited memory BFGS iterations, all full-batch, gradient-based optimization algorithm \cite[]{liu1989limited}. The L-BFGS admits smoother more robust updates
at a higher cost. We will show later the performance of both optimizers separately for comparison.

The NN functional provides a continuous representation of the wavefield, as opposed to a grid based representation, and such a continuous representation offers many benefits.  We can attain the solution at any point, no interpolation is
needed, and the domain of coverage can be of any shape. This can be beneficial in the presence of topography. However, this continuous functional representation has its limitations that appear mainly when the wavefield is complex,
requiring larger networks and more advanced training.  This appears to be the case when we have strong scattering and high frequencies. As a result, in the following tests, and as we introduce the approach, we will focus on lower frequencies 
and smoothed models. We will, nevertheless, also demonstrate these limitations.

\section{Testing the NN}

We will test this NN framework initially on two 2D examples trying to highlight some of its features and weaknesses. 
In the first example, we use a two box-shaped scatterer model with the source in the middle and we look at the dependency of the prediction on the frequency. Then, 
we apply the approach on the Marmousi model with the source on the surface, and we test the dependency of the solution on the size of the neural network. Finally, we apply the approach on a small 3D model and focus on the role of the optimizer.
 The objective of these tests is to study the ability of an NN to learn to a functional solution of the wave equation for the scattered wavefield as opposed to the wavefield itself.

\subsection{A two-scatterer model}

\begin{figure}[ht]
\begin{center}
\subfigure[]{%
\label{fig:scatter_v}
\includegraphics[width=0.3\textwidth]{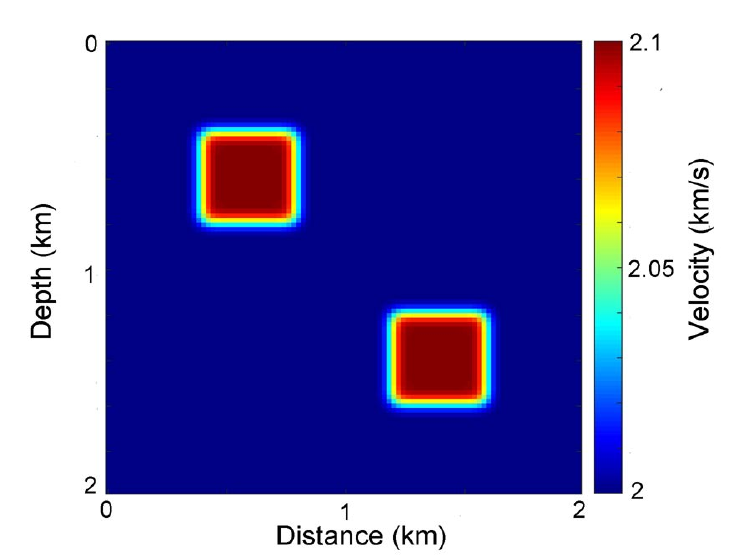}
}
\subfigure[]{%
\label{fig:u_real_5hz}
\includegraphics[width=0.3\textwidth]{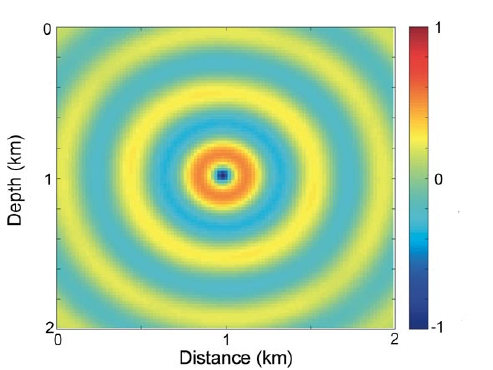}
}
\subfigure[]{%
\label{fig:u0_real_5hz}
\includegraphics[width=0.3\textwidth]{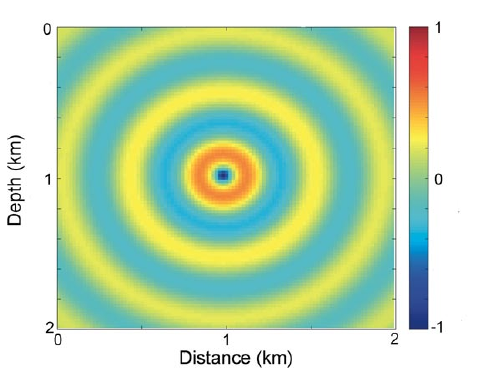}
}
\end{center}
\caption{
a) A two-scatter model. b) The real part of a 5 Hz wavefield  for the velocity in Figure~\ref{fig:scatter_v} for a source in the middle, computed numerically, and considered true. c) The real part of the 5 Hz wavefield for the background model given by a velocity of 2 km/s, computed analytically.
}%
\label{fig:scatter_v,u_real_5hz,u0_real_5hz}
\end{figure}

\begin{figure}[ht]
\begin{center}
\subfigure[]{%
\label{fig:du_real_true_5hz}
\includegraphics[width=0.3\textwidth]{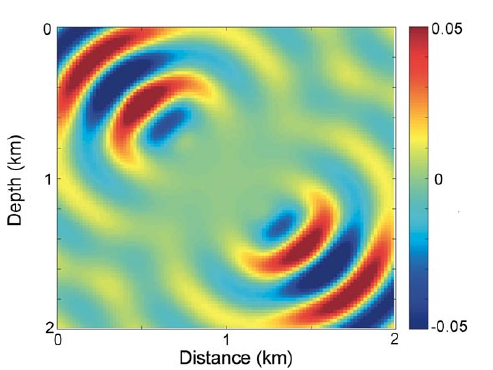}
}
\subfigure[]{%
\label{fig:du_real_ml_random5000}
\includegraphics[width=0.3\textwidth]{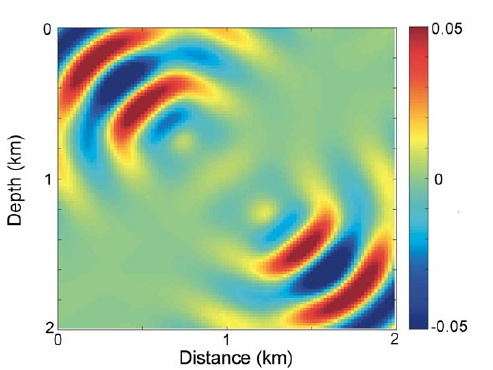}
}
\subfigure[]{%
\label{fig:du_real_dif_random5000}
\includegraphics[width=0.3\textwidth]{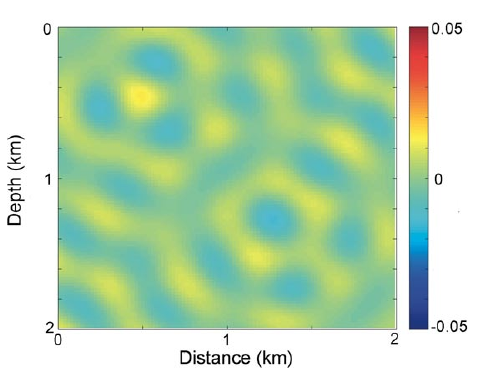}
}
\end{center}
\caption{
a) The scattered wavefield given by the difference between the two wavefields in Figures~\ref{fig:u_real_5hz} and~\ref{fig:u0_real_5hz} (True and background wavefields). b) The NN predicted scattered wavefield on a regular grid. c) The difference between the actual and predicted scattered wavefields.
}%
\label{fig:du_real_true_5hz,du_real_ml_random5000,du_real_dif_random5000}
\end{figure}

\begin{figure}[ht]
\begin{center}
\subfigure[]{%
\label{fig:u_real}
\includegraphics[width=0.4\textwidth]{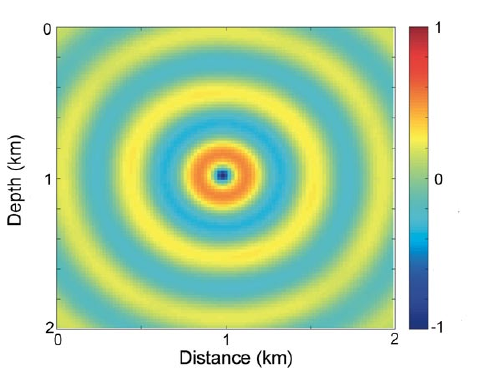}
}
\subfigure[]{%
\label{fig:u_imag}
\includegraphics[width=0.4\textwidth]{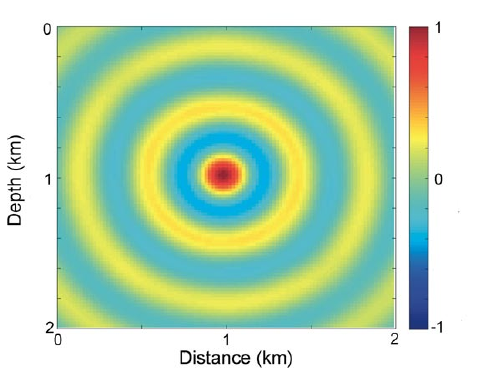}
}
\subfigure[]{%
\label{fig:u_real_ml_helmholtz}
\includegraphics[width=0.4\textwidth]{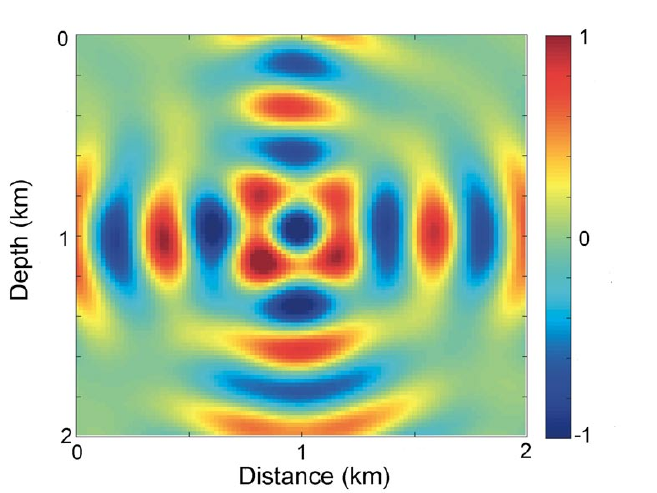}
}
\subfigure[]{%
\label{fig:u_imag_ml_helmholtz}
\includegraphics[width=0.4\textwidth]{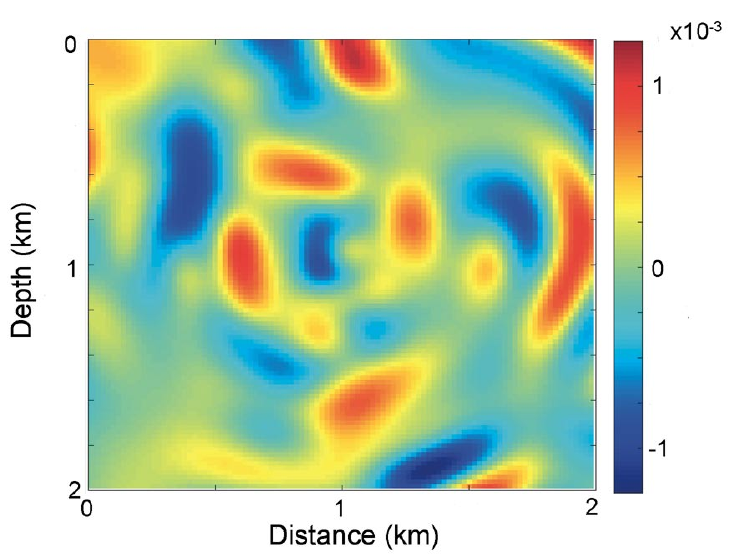}
}
\end{center}
\caption{
a) The real part of the true wavefield. b) The imaginary part of the true wavefield. c) The real part of the NN predicted wavefield. d) The imaginary part of the NN predicted wavefield. The wavefields correspond to the velocity model in Figure~\ref{fig:scatter_v}.
}%
\label{fig:u_real,u_imag,u_real_ml_helmholtz,u_imag_ml_helmholtz}
\end{figure}

In the first model, we place two box-shaped perturbations in an otherwise homogeneous background as shown in Figure~\ref{fig:scatter_v}. The model has 100 samples in both the $x$ and $z$ directions,
with a sampling interval of 20 m. The corresponding (real part) of the 5 Hz wavefield for a point source (a $delta$ function, one sample) in the center of the model
is shown in Figure~\ref{fig:u_real_5hz}. The background model is given by a constant velocity of 2 km/s, and the corresponding wavefield for the same source and frequency is shown in Figure~\ref{fig:u0_real_5hz}. 
If we subtract the two wavefields, we obtain the true scattered wavefield with the real part shown in Figure~\ref{fig:du_real_true_5hz}, where the energy, as expected, reflects scattering from the two box-shaped scatterers. 
Using the loss function in equation~\ref{eqn:eq3}, we train an 8-layer deep fully connected neural network with 20 neurons in each layer to represent the scattered wavefield solution. We randomly chose 5000 samples from the space domain ($x_i,z_i$) for the training, and train for 100000 epochs of Adam updates and 20000 of LBFGS updates. This number of samples used 
represents one fourth of the grid samples used to solve the Helmholtz equation and it was necessary to arrive to the scattered
wavefield solution shown in Figure~\ref{fig:du_real_ml_random5000}. The difference between the true scattered wavefield and the NN predicted one is shown in Figure~\ref{fig:du_real_dif_random5000}. There are differences, but
they are generally mild. The imaginary part of the scattered wavefield, not shown here, had similar accuracy.

\begin{figure}[ht]
\begin{center}
\subfigure[]{%
\label{fig:du_real_ml8hz}
\includegraphics[width=0.3\textwidth]{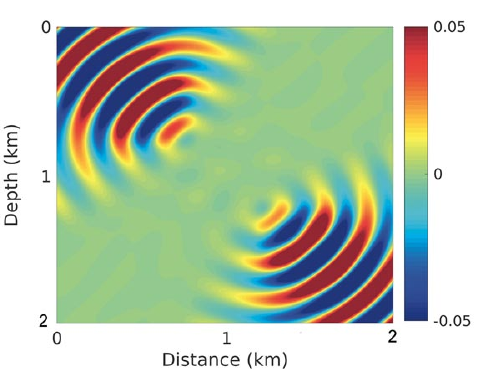}
}
\subfigure[]{%
\label{fig:du_real_dif8hz}
\includegraphics[width=0.3\textwidth]{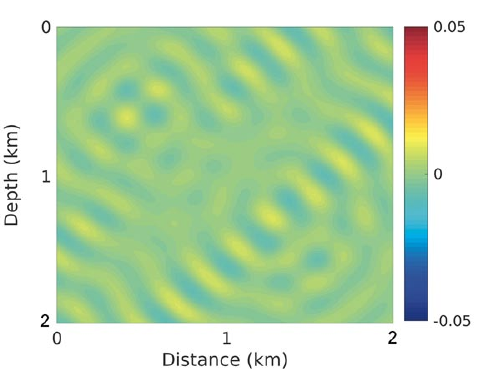}
}
\subfigure[]{%
\label{fig:misfit_8hz}
\includegraphics[width=0.3\textwidth]{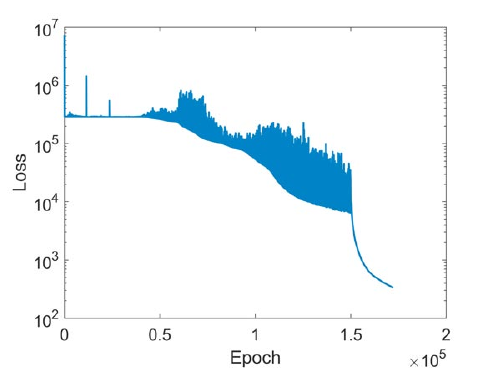}
}
\end{center}
\caption{
a) The NN predicted scattered 8Hz wavefield for a source in the middle. b) The difference between the predicted scattered wavefield and the one computed numerically (true) plotted at the same scale as in a). c) The NN training loss function, which displays the loss using Adam followed by LBFGS.
}%
\label{fig:du_real_ml8hz,du_real_dif8hz,misfit_8hz}
\end{figure}
 
To justify inverting for the scattered wavefield instead of the wavefield directly using the Helmholtz wave equation, we repeat the exact experiment with the same number of randomly chosen training samples.
The loss function, in this case, is given by the Helmholtz wave equation and to lessen the effect of a point source bias, we use an isotropic Gaussian source with a variance of 2.5. Figures~\ref{fig:u_real} and~\ref{fig:u_imag} show the real and imaginary parts, respectively, of
the true wavefield for the two-scatterers model shown Figure~\ref{fig:scatter_v}. Figures~\ref{fig:u_real_ml_helmholtz} and~\ref{fig:u_imag_ml_helmholtz} show the real and imaginary parts, respectively, of
the NN predicted wavefield for the same model. The difference is large and this is attributed to the source singularity in the Helmholtz equation, which requires better sampling of the source area in the training data.  

For an 8 Hz wavefield, we use a larger network given by 40 neurons in each of the 8 layers and we use 10000 samples in the training. 
The real part of the predicted scattered wavefield is shown in Figure~\ref{fig:du_real_ml8hz}. The difference between this predicted wavefield and the 
considered true numerical solution, plotted at the same scale as in Figure~\ref{fig:du_real_ml8hz}, is small as shown in Figure~\ref{fig:du_real_dif8hz}. To arrive to this solution, we used 150000 epochs of Adam updates and
20000 of LBFGS updates as demonstrated in Figure~\ref{fig:misfit_8hz}. We use LBFGS at the end as it admits smoother updates we can rely on, but it is generally more expensive. The 
sudden change in the behaviour of the loss curve reflects the transition from Adam to LBFGS.
Despite the larger network, compared to the 5Hz case, and additional epochs, the cost increase was less than $100\%$, and that is much smaller than the additional cost we experience in solving for high frequencies using finite difference methods, which tend to increase exponentially.

\subsection{The Marmousi model}

\begin{figure}[ht]
\begin{center}
\subfigure[]{%
\label{fig:vM}
\includegraphics[width=0.3\textwidth]{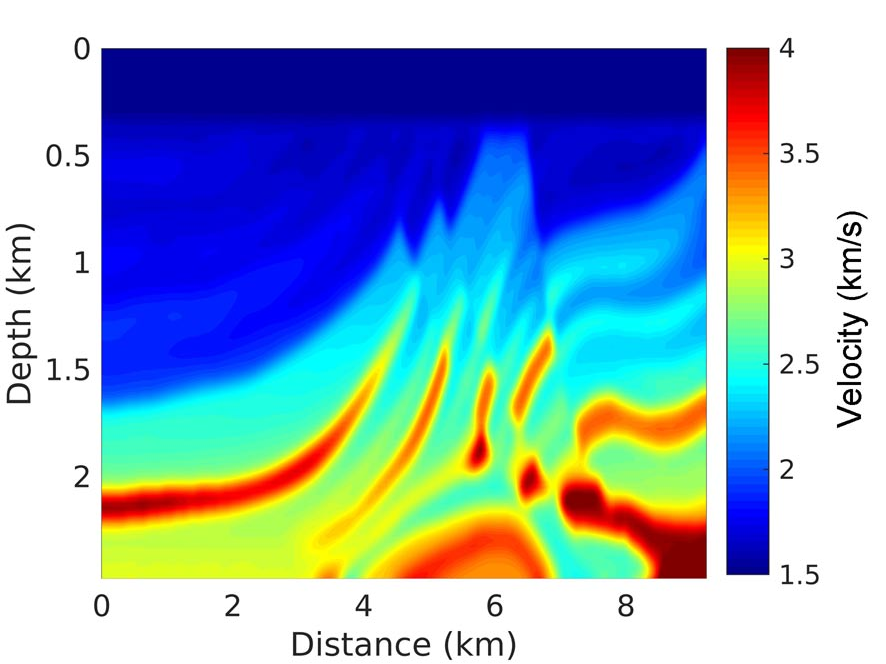}
}
\subfigure[]{%
\label{fig:du_real_true}
\includegraphics[width=0.3\textwidth]{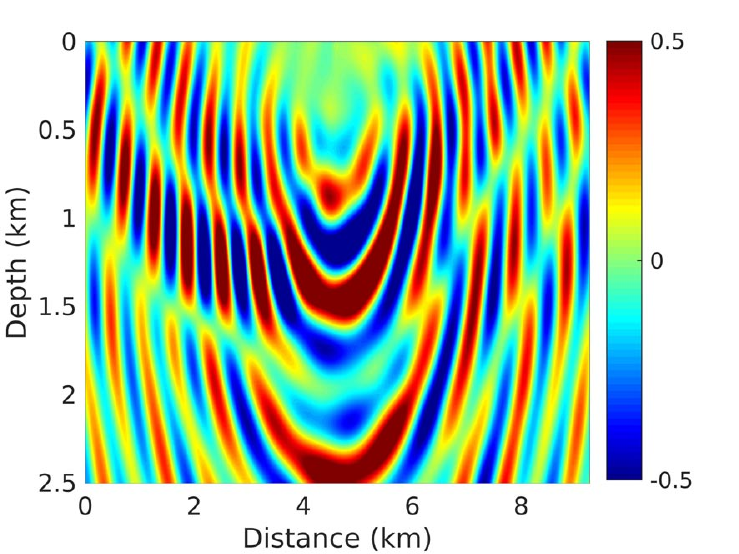}
}
\subfigure[]{%
\label{fig:du_imag_true}
\includegraphics[width=0.3\textwidth]{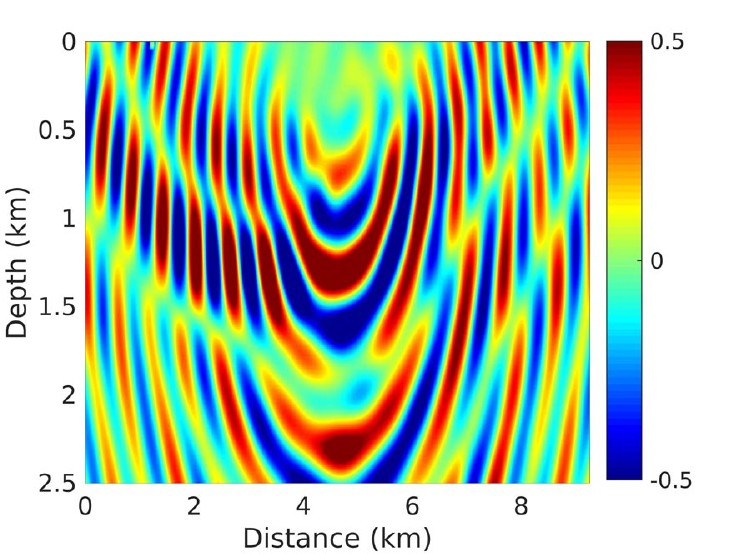}
}
\end{center}
\caption{
a) The Marmousi model. b) The real part of the resulting 3 Hz wavefield for a source on the surface located in the middle. c) The imaginary part of the wavefield.
}%
\label{fig:vM,du_real_true,du_imag_true}
\end{figure}

\begin{figure}[ht]
\begin{center}
\subfigure[]{%
\label{fig:misfit}
\includegraphics[width=0.3\textwidth]{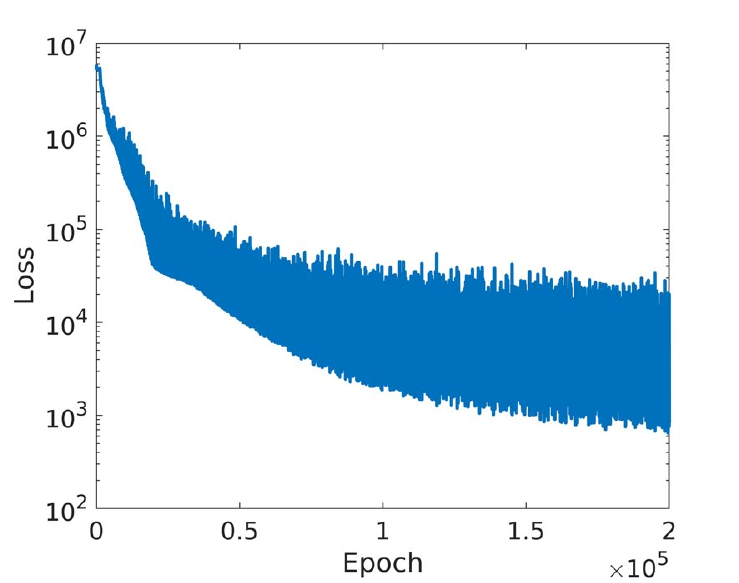}
}
\subfigure[]{%
\label{fig:du_real_ml}
\includegraphics[width=0.3\textwidth]{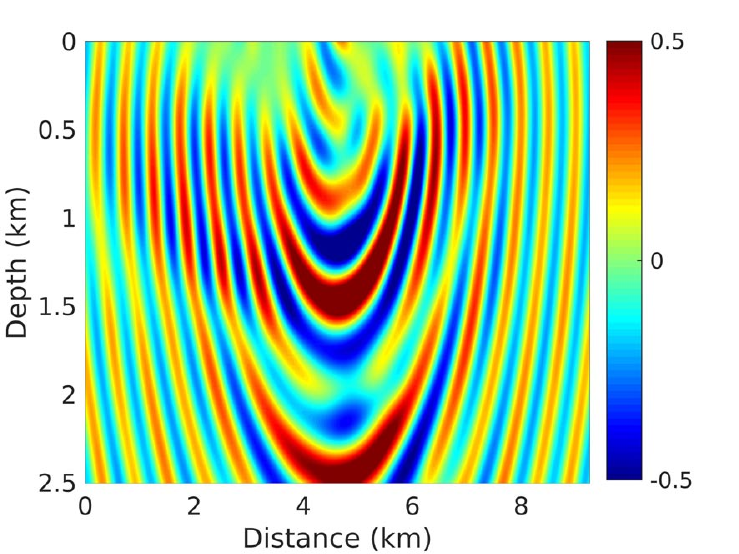}
}
\subfigure[]{%
\label{fig:du_imag_ml}
\includegraphics[width=0.3\textwidth]{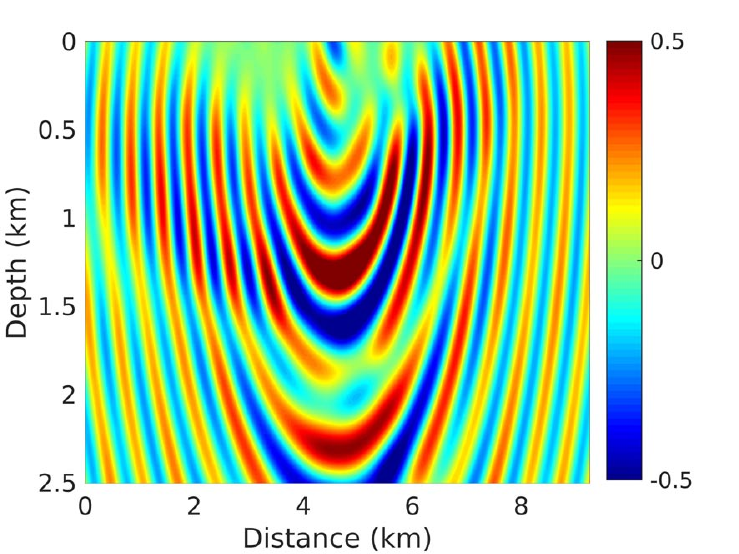}
}
\end{center}
\caption{
a) The loss function for the training of the NN. b) The real part of the predicted scattered wavefield from the NN network. c) The imaginary part.
}%
\label{fig:misfit,du_real_ml,du_imag_ml}
\end{figure}

Now, we test the utilization of the NN PDE in solving for the wavefield for a slightly smoothed Marmousi model (Figure~\ref{fig:vM}). A point source is placed, this time, on the surface at location 4.5 km.
 We solve the Helmholtz equation numerically to obtain the 3 Hz frequency wavefield. 
The background model is homogeneous with a velocity of 1.5 km/s in which we can solve the wavefield analytically. The difference between the true and the background wavefields, constituting the scattered wavefield 
is shown in Figures~\ref{fig:du_real_true} (real part) and~\ref{fig:du_imag_true} (imaginary part). The background wavefield and the model perturbations (difference between the true model and the homogeneous background)
are used in the cost function given by equation~\ref{eqn:eq3} to invert for the NN parameters. We use, this time, a 10-layer network with \{128, 128, 64, 64, 32, 32, 16, 16, 8, 8\} neurons in the layers, respectively.
We find that this configuration, given by larger dimensional layers early, is generally more effective. We use 10000 random sample points for the training and the resulting loss over 20000 epochs of training is shown in Figure~\ref{fig:misfit}. 
The trained network is then used to evaluate the scattered wavefield on a regular grid and the resulting real part is shown
in Figure~\ref{fig:du_real_ml} and the imaginary part is shown in Figure~\ref{fig:du_real_ml}. 

\begin{figure}[ht]
\begin{center}
\subfigure[]{%
\label{fig:du_real_dif}
\includegraphics[width=0.3\textwidth]{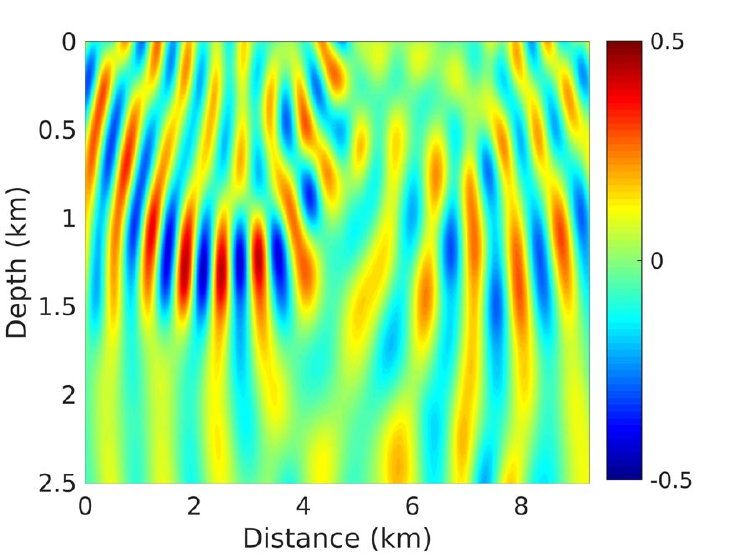}
}
\subfigure[]{%
\label{fig:du_imag_dif}
\includegraphics[width=0.3\textwidth]{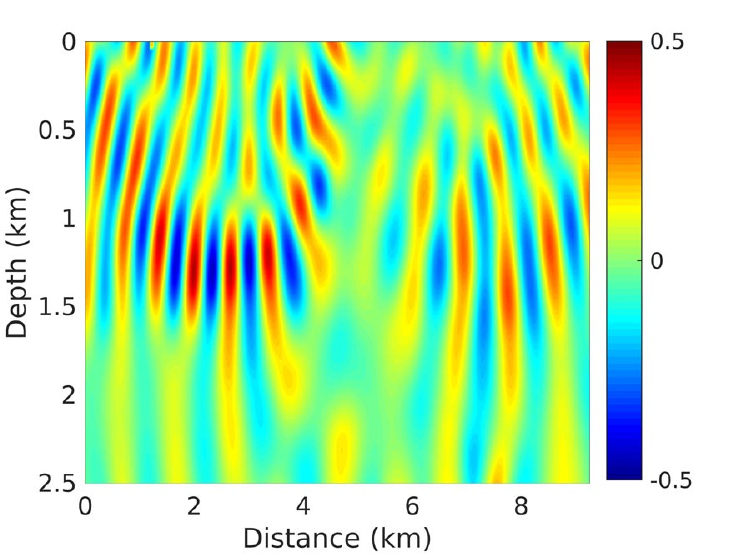}
}
\subfigure[]{%
\label{fig:v0}
\includegraphics[width=0.3\textwidth]{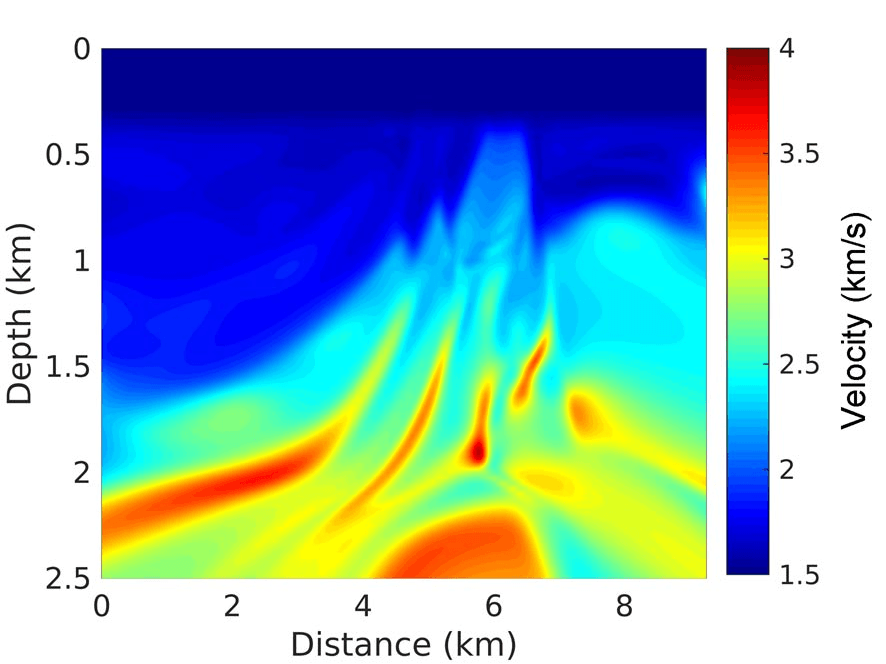}
}
\end{center}
\caption{
a) The difference between Figures~\ref{fig:du_real_true} and~\ref{fig:du_real_ml} (True and predicted real parts of the scattered wavefields). b) The difference between Figures~\ref{fig:du_imag_true} and~\ref{fig:du_imag_ml} (True and predicted imaginary parts of the scattered wavefields). c) The velocity model computed from the predicted wavefield.
}%
\label{fig:du_real_dif,du_imag_dif,v0}
\end{figure}

\begin{figure}[ht]
\begin{center}
\subfigure[]{%
\label{fig:du_real_ml_n64}
\includegraphics[width=0.3\textwidth]{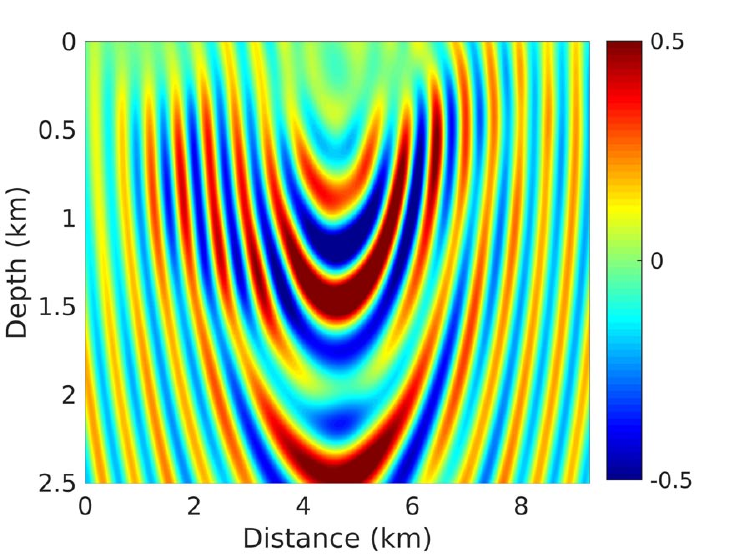}
}
\subfigure[]{%
\label{fig:du_real_dif_n64}
\includegraphics[width=0.3\textwidth]{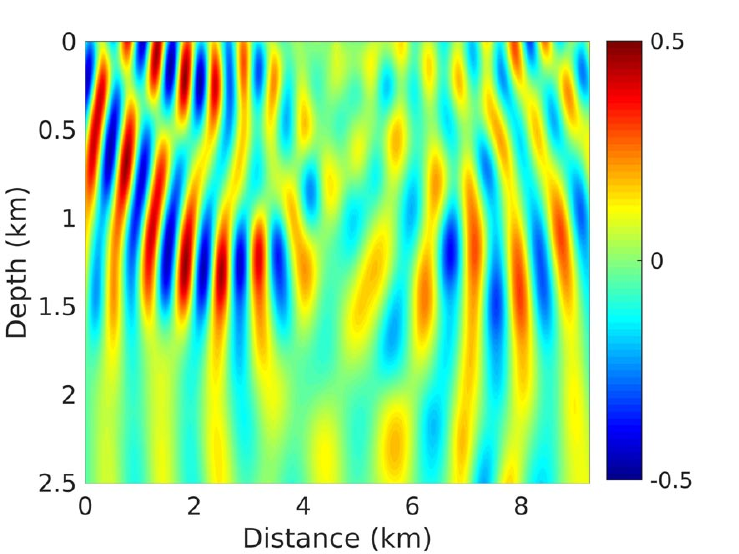}
}
\subfigure[]{%
\label{fig:v_pred_n64}
\includegraphics[width=0.3\textwidth]{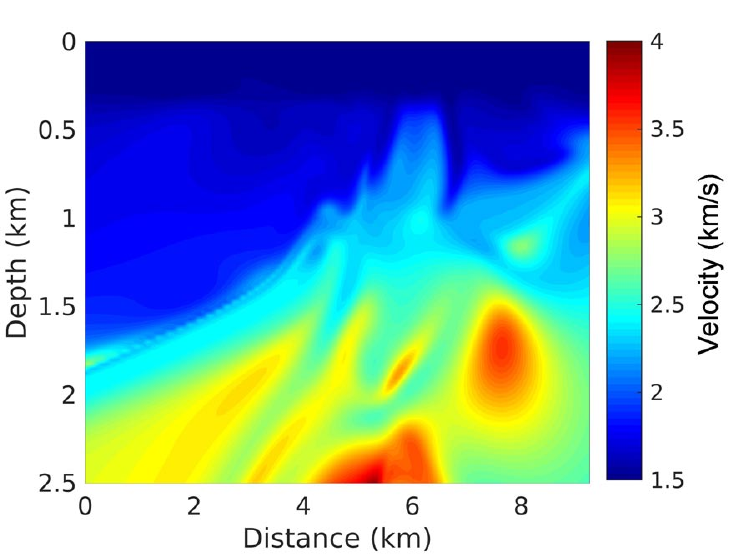}
}
\end{center}
\caption{
a) The real part of the predicted scattered wavefield using an 8-layer network. b) The difference between Figures~\ref{fig:du_real_true} and~\ref{fig:du_real_ml_n64} (True and predicted real parts of the scattered wavefields). c) The velocity model computed from the predicted wavefield.
}%
\label{fig:du_real_ml_n64,du_real_dif_n64,v_pred_n64}
\end{figure}

\begin{figure}[ht]
\begin{center}
\subfigure[]{%
\label{fig:du_real_ml_n256}
\includegraphics[width=0.3\textwidth]{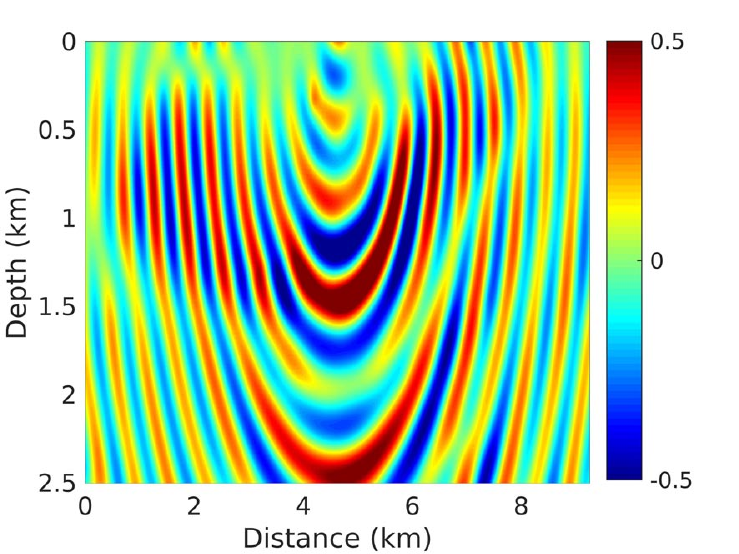}
}
\subfigure[]{%
\label{fig:du_real_dif_n256}
\includegraphics[width=0.3\textwidth]{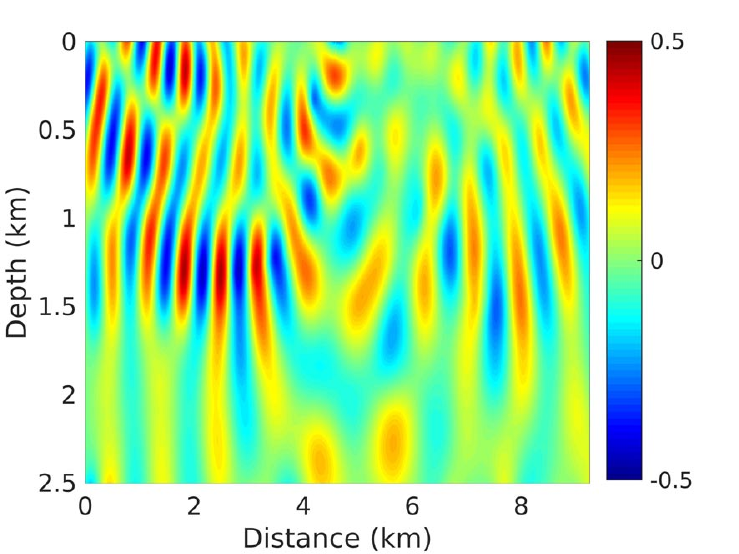}
}
\subfigure[]{%
\label{fig:v_pred_n256}
\includegraphics[width=0.3\textwidth]{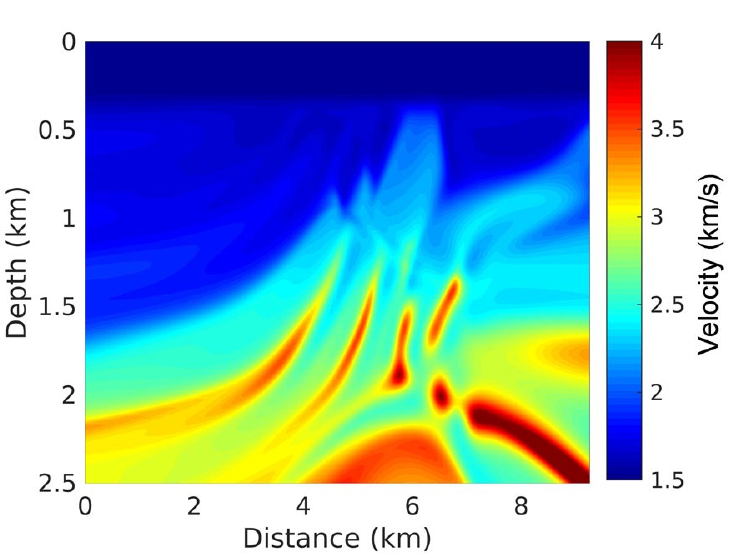}
}
\end{center}
\caption{
a) The real part of the predicted scattered wavefield using a 12-layer network. b) The difference between Figures~\ref{fig:du_real_true} and~\ref{fig:du_real_ml_n256} (True and predicted real parts of the scattered wavefields). c) The velocity model computed from the predicted wavefield.
}%
\label{fig:du_real_ml_n256,du_real_dif_n256,v_pred_n256}
\end{figure}

The difference between the true scattered wavefield and the NN predicted one is shown in Figures~\ref{fig:du_real_dif} (real part)
and~\ref{fig:du_imag_dif} (imaginary part). The difference is generally small again, but here it seems to include more coherent energy corresponding to some of the scattering. In other words, the resulting NN predicted scattered
wavefield is smoother than the true wavefield. This is an expected feature of NN when we avoid overfitting, the network acts as a smoother \cite{neal2018modern}. We can further verify this smoothness feature by using the wave equation to compute the velocity model corresponding to the predicted wavefield as shown in Figure~\ref{fig:v0}.

If we use a smaller network of 8 layers with \{64, 64, 32, 32, 16, 16, 8, 8\} neurons in the layers from left to right (we dropped the first two layers from the previous network), and use the same number of epochs, the resulting real part of the predicted scattered wavefield tends to be smoother as shown in Figure~\ref{fig:du_real_ml_n64}. The difference between the predicted and true scattered wavefield is shown in Figure~\ref{fig:du_real_dif_n64} plotted at the same scale. The difference includes more energy than before. 
The increased smoothness of the wavefield can be verified by the resulting velocity model calculated from the wavefield and shown in Figure~\ref{fig:v_pred_n64}. The velocity model is smooth compared to the true model,
reflecting the smooth nature of the wavefield.

On the other hand, if we actually use a 12-layer network by adding two layers at the beginning of the original network with 256 neurons in each of them, we end up with a network given by
\{256, 256, 128, 128, 64, 64, 32, 32, 16, 16, 8, 8\} neurons in the layers from left to right. Using the same number of epochs in the training of the same random samples in space, we end up with 
the predicted scattered wavefield with the real part shown in Figure~\ref{fig:du_real_ml_n256}. The difference between the predicted and true scattered wavefield is shown in Figure~\ref{fig:du_real_dif_n256} plotted at the same scale.
It contains less energy and that again can be verified by the resulting velocity model calculated from the wavefield and shown in Figure~\ref{fig:v_pred_n256}.
The velocity model is clearly sharper and it is reasonably close to the true velocity model shown in Figure~\ref{fig:vM}. 

\begin{figure}[ht]
\begin{center}
\subfigure[]{%
\label{fig:v3D}
\includegraphics[width=0.3\textwidth]{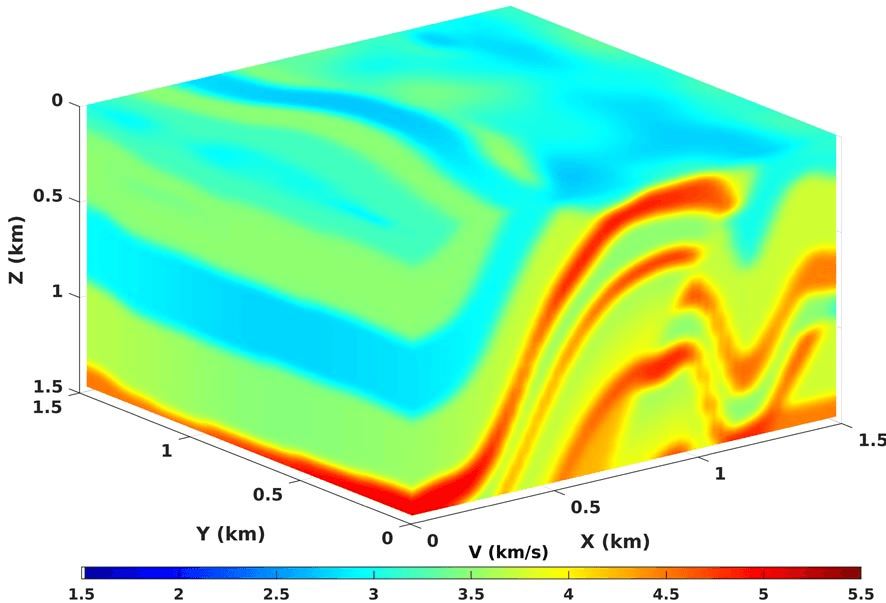}
}
\subfigure[]{%
\label{fig:du1_real_true3D}
\includegraphics[width=0.3\textwidth]{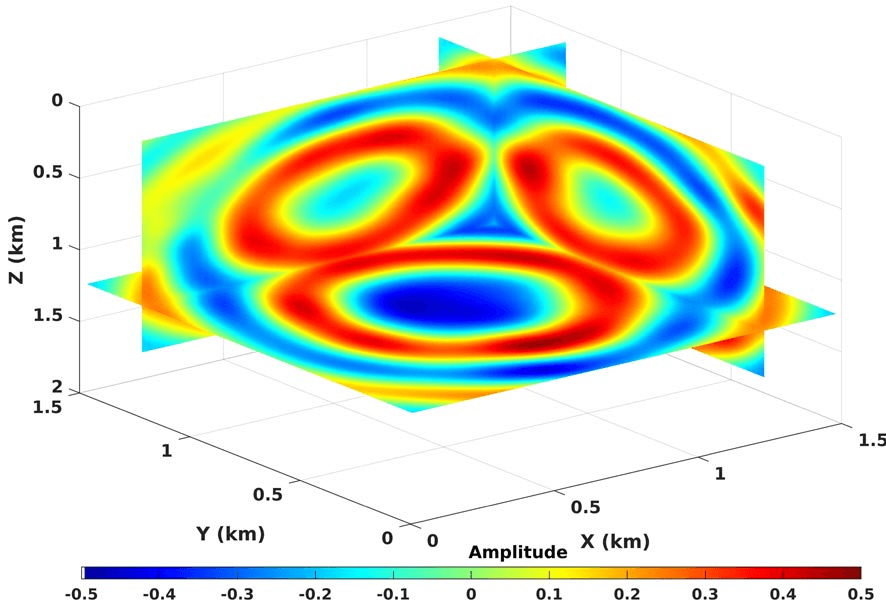}
}
\subfigure[]{%
\label{fig:DNN_3d}
\includegraphics[width=0.3\textwidth]{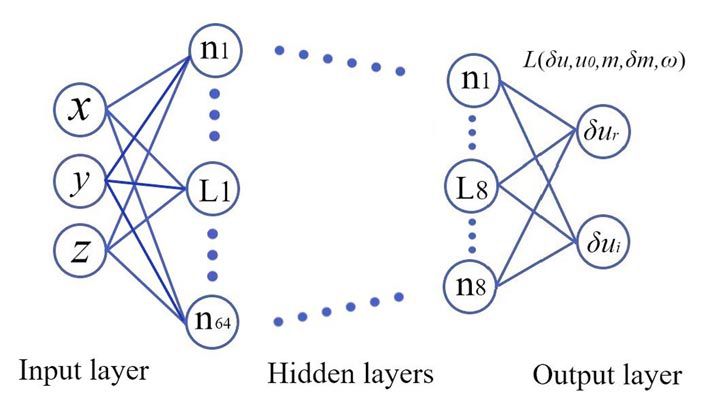}
}
\end{center}
\caption{
a) A 3D model. b) The real part of the resulting 10 Hz wavefield for a source on the surface located in the middle. c) The NN architecture with dimensions of the 8 hidden layers given by (64,64,32,32,16,16,8,8) from shallow to deep.
}%
\label{fig:v3D,du1_real_true3D,DNN_3d}
\end{figure}	
 
Thus, a larger network can provide more accurate wavefields, but for applications in gradient calculation for velocity model update, a perfect scattered wavefield is not necessary. The cost of training the 
12-layer network is 50$\%$ higher than the 10-layer one in spite that the number of network model parameters increased by 4. Meanwhile, the cost of the training the 8-layer network is two-third the cost of training the 10-layer network,
while the number of network parameters is one-fourth of that of the 10-layer network. So, in summary, using this network architecture, the cost of the training will increase by about 50$\%$ with the addition of two layers of double the
size of the first (largest) layer, and we end up with a higher resolution wavefield.

\subsection{A 3D example and the optimizer}

We consider a 3D cube extracted from the SEG/EAGE Overthrust model \cite{aminzadeh1994seg} and slightly smoothed as shown in Figure~\ref{fig:v3D}. In this test, we also test the performance of two NN optimization algorithms, specifically Adam and LBFGS. The background homogeneous model has a velocity of 3.2 km/s. The difference between the Helmholtz computed
wavefield for 10 Hz and the background wavefield for the same frequency provides us to the true scattered wavefield for a source located in the middle, with the real part of this scattered wavefield shown in Figure~\ref{fig:du1_real_true3D}.
Since the Adam optimizer admits, as we saw earlier, turbulent loss functions,
for this example, we will test the performance of the two network optimization algorithms separately: The Adam
optimizer and the limited memory BFGS algorithm. In this comparison, for the Adam optimizer we use 150000 epochs and for the LBFGS we use 50000 epochs in the training, and the neurons in each of the 8 hidden layers are $\{64,64,32,32,16,16,8,8\}$, as shown in Figure~\ref{fig:DNN_3d}. 

\begin{figure}[ht]
\begin{center}
\subfigure[]{%
\label{fig:misfit_over_adam64}
\includegraphics[width=0.3\textwidth]{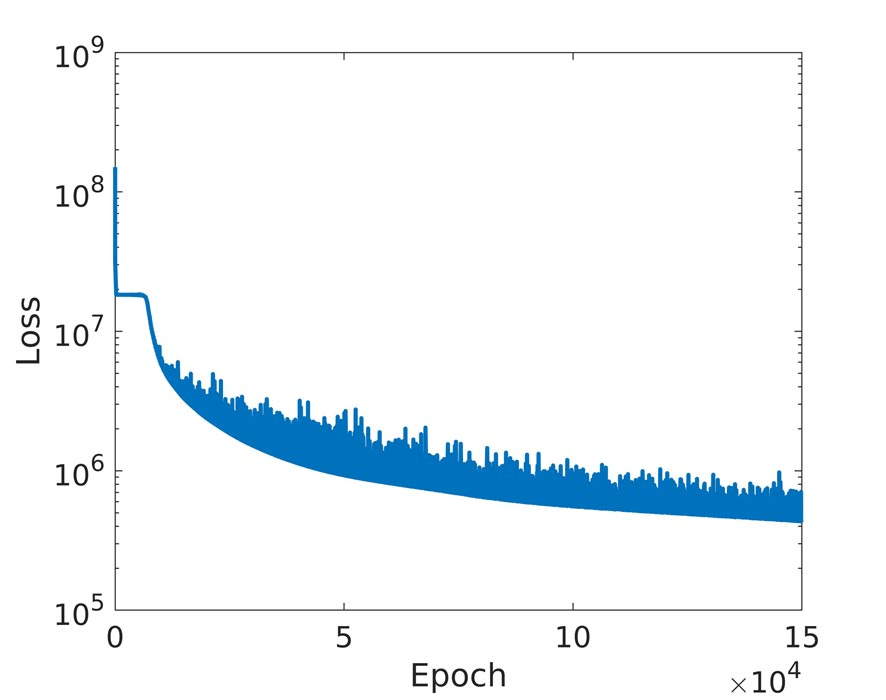}
}
\subfigure[]{%
\label{fig:du1_real_ml_adam}
\includegraphics[width=0.3\textwidth]{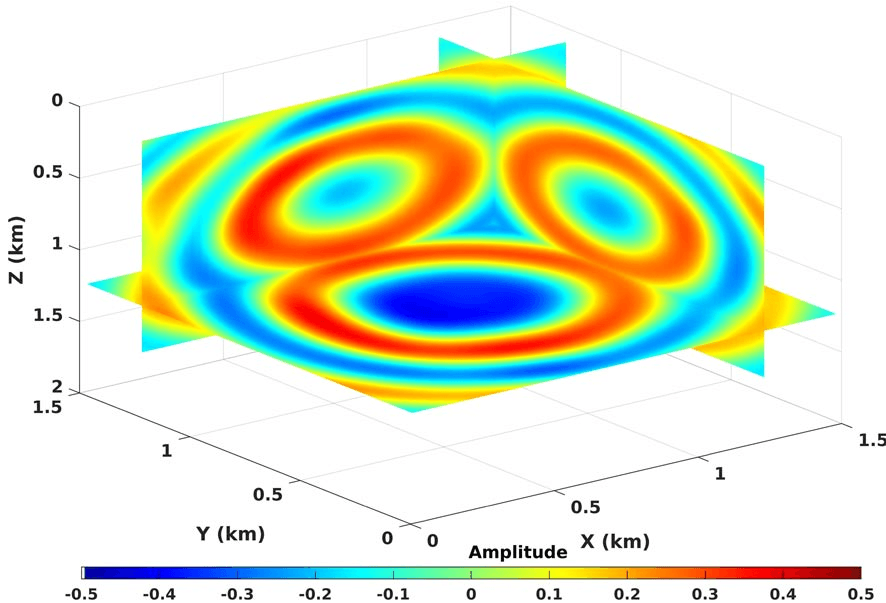}
}
\subfigure[]{%
\label{fig:du1_real_dif_adam}
\includegraphics[width=0.3\textwidth]{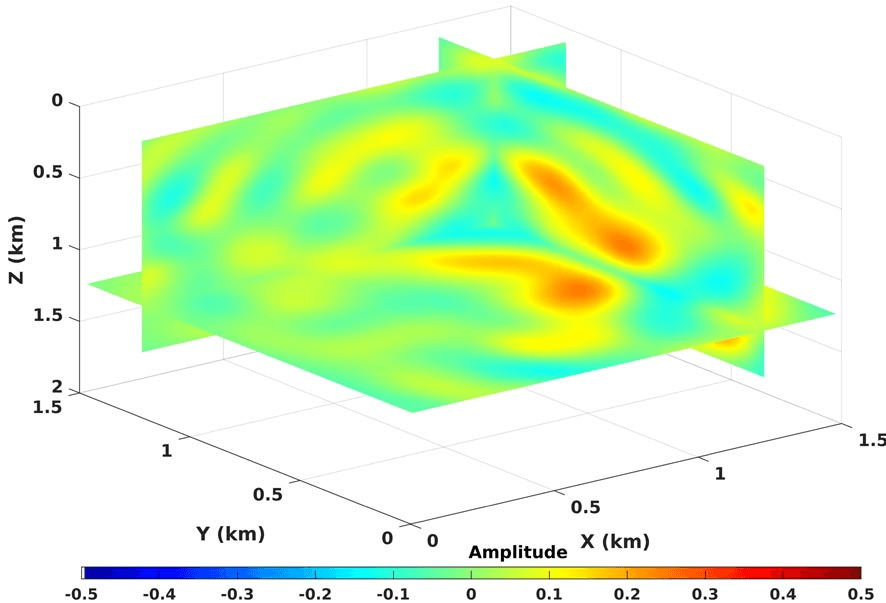}
}
\end{center}
\caption{
a) The loss function for the training of the NN using an Adam optimizer. b) The real part of the predicted scattered wavefield from the NN. c) The difference between the predicted wavefield and the true one in Figure~\ref{fig:du1_real_true3D}.
}%
\label{fig:misfit_over_adam64,du1_real_ml_adam,du1_real_dif_adam}
\end{figure}		

\begin{figure}[ht]
\begin{center}
\subfigure[]{%
\label{fig:misfit_over_lbfgs}
\includegraphics[width=0.3\textwidth]{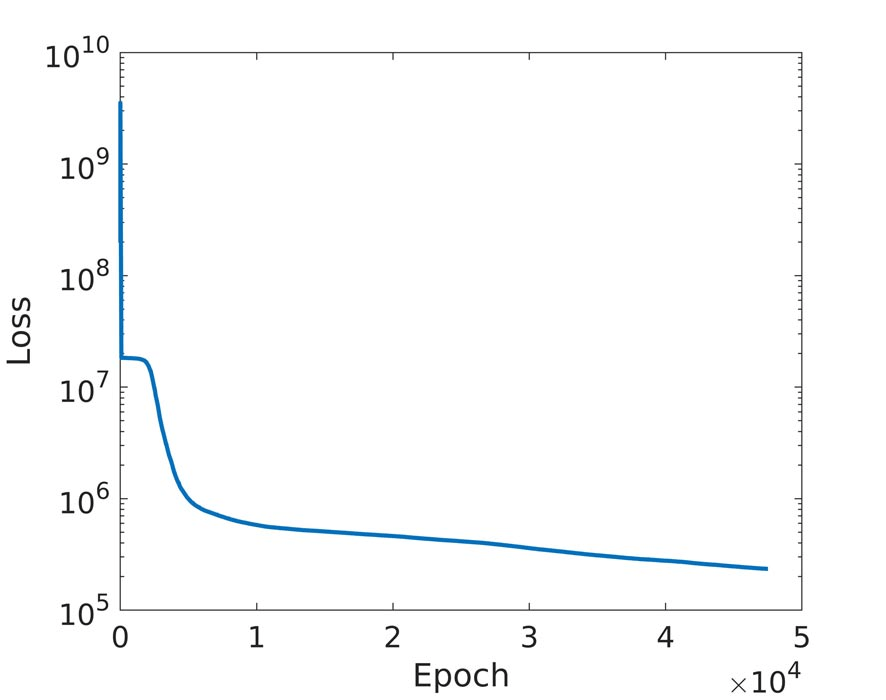}
}
\subfigure[]{%
\label{fig:du1_real_ml_lbfgs2020}
\includegraphics[width=0.3\textwidth]{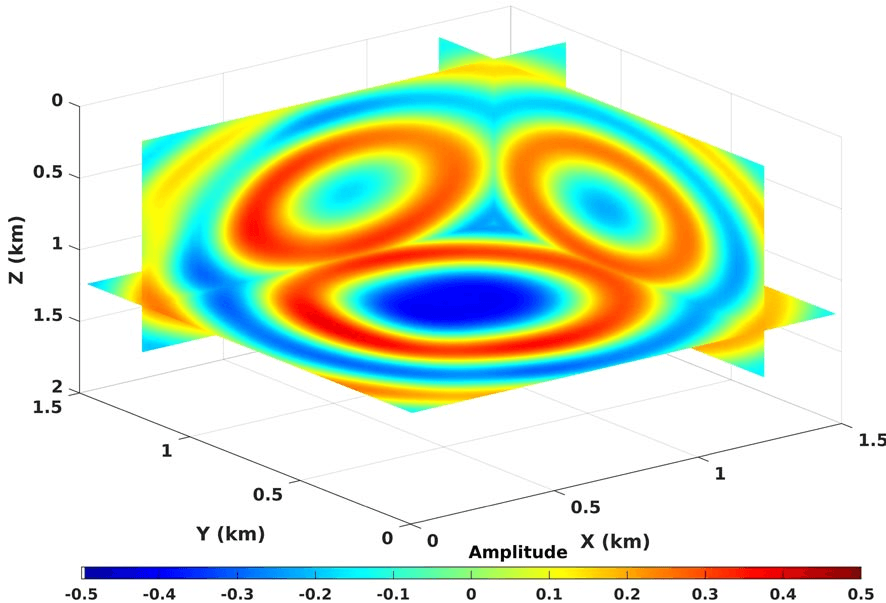}
}
\subfigure[]{%
\label{fig:du1_real_dif_lbfgs2020}
\includegraphics[width=0.3\textwidth]{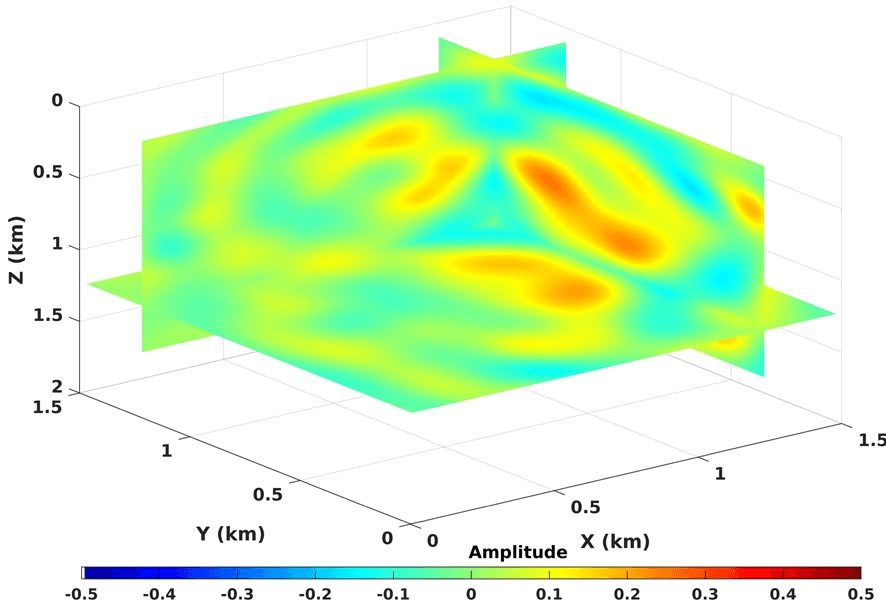}
}
\end{center}
\caption{
a) The loss function for the training of the NN using an LBFGS optimizer. b) The real part of the predicted scattered wavefield from the NN. c) The difference between the predicted scattered wavefield and the true one in Figure~\ref{fig:du1_real_true3D}.
}%
\label{fig:misfit_over_lbfgs,du1_real_ml_lbfgs2020,du1_real_dif_lbfgs2020}
\end{figure}

The loss function for the Adam optimizer is shown in Figure~\ref{fig:misfit_over_adam64} and in average the loss reduces per epoch, but with Adam we notice the loss function is bumpy and this has been realized by 
others in the application of PINN. Considering the log scale of the vertical axis, such alterations are slightly exaggerated in the Figure. The Adam update can be made smoother by using a smaller learning rate, but that increases the number of epochs. The real part of the resulting predicted scattered wavefield is shown in Figure~\ref{fig:du1_real_ml_adam}, and the
difference between it and the true scattered wavefield is shown in Figure~\ref{fig:du1_real_dif_adam}. The difference, plotted at the same scale as the scattered wavefield, is small. On the other hand, using an LBFGS optimizer,
the loss function is smoother as shown in Figure~\ref{fig:misfit_over_lbfgs}, and seemingly admits a lower loss. However, the predicted scattered wavefield from the network, shown in Figure~\ref{fig:du1_real_ml_lbfgs2020}, looks 
almost identical to the one obtained from the Adam optimizer. This can be further verified by observing the difference between the predicted scattered wavefield and the true one shown in Figure~\ref{fig:du1_real_dif_lbfgs2020},  plotted again at the same scale. The errors seem to be similar to that observed with the Adam optimizer. So the effective differences in the loss has limited effect on the wavefield. 

Considering that the results from using the Adam and the LBFGS optimizers are similar, and since the LBFGS optimizer is more expensive depending on the memory parameters, we suggest, as \cite{raissi2019physics} suggested, using initially the Adam optimizer, which is extremely popular in ML
optimizations. 

\section{Discussions}

Machine learning provides a platform for predicting outputs by mainly recognizing the corresponding patterns of the inputs through a training process. Wavefields are by definition smooth and differentiable other than at the source,
which is a  requirement for a functional NN solution output \cite{hornik1989multilayer}.
The input to the proposed neural network is a location in space and the output is the wavefield (or the scattered wavefield) that satisfies a cost function given by the Helmholtz  equation or a variation of it. These equations depend on the velocity and the source function, or in our implementation, the background wavefield
and the velocity perturbations. So the NN weights and biases are expected to 
absorb the velocity and source information in their efforts to learn to predict the solution of the wave equation. As the NN tries to learn the wavefield, the source location is, especially, influential as it determines the epicentre of the wavefield. The velocity has generally a second-order effect on the wave shape, compared to the source location. 
Meanwhile, the frequency mainly controls the wavelength. These facts can help us decide on how to use the network for any successive wavefield solutions. For example, solutions for any additional velocity perturbations that maybe extracted from any velocity model 
update procedure like migration velocity analysis or full waveform inversion. Specifically, the current NN model can be used as an initial model for training on the updated velocity.

In using the Born (Lippmann–Schwinger equation) version of the wave equation instead of the Helmholtz solver, we avoided the bias required in better sampling the source region in the training of the network, necessary to mitigate the effect of the source singularity. 
So by using a homogeneous background model in which the wavefield can be solved analytically (and instantly), the perturbations (difference between true and background models) will often extend the model domain,
and random samples of the model space can be used in the training. To remove the signature of the source singularity from the scattered wavefield, the background velocity should be chosen equal to the velocity at the source. 
For marine data, this is given by a velocity of approximately 1.5 km/s. In general, the accuracy of predicting the wavefield depends mainly on the complexity of the wavefield. Near the source, the wavefield is complex, but it could also be complex
in many other areas depending on the velocity model. In this case, we will need a larger neural network model, as well as better sampling of these complex regions. The random training points will often sample the domain reasonably well, but
it does not take into account the complexity of the wavefield. From our observation, and especially with the Marmousi model, for a fixed neural network model size and random sampling of the training, the NN provides a uniformly smooth
wavefield compared the true one.
We can also utilize the concept of collocation points, as well, as adaptively adding points in regions requiring 
more emphasis (i.e., with high residuals) \cite{mcfall2009artificial}. These options have their own cost. The number of training samples used to train the NN to predict the scattered wavefield is a delicate matter \cite{zhou2011analyses}.  It directly affects the cost of the training and yet it is necessary that we have enough samples to accurately train the network to predict the wavefield. Training examples and their influence on the training is an ongoing research topic in the machine learning community.

Another feature of using a cost function for NN training, like in \cite{raissi2019physics}, we can fit the boundary condition or even the data as part of the objective, and thus, include two or more terms in the cost function. 
For the data fitting case, this amounts to something like the wavefield reconstruction method \cite{van2013mitigating}, which is also
solved in the frequency domain and faces similar challenges with regard to data and model sizes \cite{song2021solving}. Thus, an important feature of such neural network wavefield solutions is the fixed memory requirements, mainly controlled by the architecture of the network. It is, thus, independent of the size of the gridded velocity model. As we saw, the errors associated with reducing the size of the network are not of the dispersion kind, like for 
conventional numerical solvers considering the velocity model discretization, but they manifest themselves in smoothing the wavefield.

The cost of training the neural network depends on the number of random samples used in the training (the training set) to optimize the network parameters, as well as, the size of the network.
As the trained NN tries to fit the loss function given by the wave equation or its Born form and any boundary conditions, the size of the network, including the number of layers and neurons, defines the details of the
predicted wavefield. As we saw, smaller networks,  cheaper to train, admit smoother wavefields. Thus, the size of the neural network will depend on the application and the objective involved in the application.
An application like waveform inversion may require smoother wavefields in the early iterations as we build up the background low wavenumber model, and thus, a neural network wavefield can help us obtain 
smoother velocities and/or gradients without the need for smoothing or spatial filtering. 
Intuitively, the smoother the wavefield the less parameters we will need, which bodes well for low frequencies. Interestingly, the cost increase of enlarging the network to predict wavefields for higher frequencies, is less severe
than that needed for finite difference methods. However, we noticed that for high frequencies the training is harder as we use inverse tangent activation functions to develop the sinusoidal wavefield. An
alternative is to use sine activation functions, which we plan to investigate in the near future.

Though, at this early stage of using this functional ML solution, the cost may not justify replacing the regular Helmholtz equation for 2D, and maybe even 3D, isotropic examples, the potential and flexibility 
of the approach will induce more interesting applications. This includes applications on more complex physics, like anisotropy and elasticity, where the regular Helmholtz solver becomes impractical. 
This includes applications involving complex wavefields in 3D, like those for orthorhombic anisotropy, even if the perturbations are small.
Conventional frequency-domain solutions for such complex physics are hard and somewhat beyond our capability, especially if the model size is large. Machine learning is an optimal platform for large problems and  large data as it adapts to this objective and
learns the appropriate solution by recognizing patterns. Thus, for complex physics, our cost function will change, and possibly include more terms, but the training machinery is the same and does not involve solving for the inverse of a large matrix.

\section{Conclusions}

We trained a neural network to provide functional solutions to
the Helmholtz equation. To avoid the point source singularity, we use a fully connected network that takes in space coordinates within the domain of interest and 
outputs the real and imaginary parts of the scattered (instead of the full) wavefield in the frequency domain. The background velocity is
homogeneous, which admits analytical solutions of the background wavefield. With automatic differentiation, the network is capable, as well, of evaluating the partial derivatives of the scattered wavefield necessary to evaluate the loss function given by the Lippmann Schwinger form of the wave equation. This loss function is used to update the network parameters.
With a scattered wavefield corresponding to perturbations spanning the space domain, the training of the network can be performed with less random samples. However, the network and the number of samples should
increase with an increase in frequency. This increase is far less than the exponential increase we experience in the case of increasing frequency for finite difference methods. 
Overall, the output wavefields are somewhat smoother than the exact ones,
and this can be attributed to the compromise feature of our relatively small network and this feature might be useful for applications like waveform inversion. In fact, the smaller the network, the smoother the output
scattered wavefield.

\newpage


%% References



%% New version of the num-names style
\bibliographystyle{elsarticle-num-names}
\bibliography{pinn_helmholtz}
%\biboptions{comma}

\end{document}